\documentclass[preprint2]{aastex63}

\usepackage{url}
\usepackage[normalem]{ulem}
\received{March 27, 2020}
\revised{May 6, 2020}
\accepted{May 27, 2020}
\submitjournal{ApJL}

\def\fv{ }

\newcommand{\frbp}{FRB~180916 }
\newcommand{\frb}{FRB~180916}
\def\agile{{\fv{\it AGILE}} }
\def\agilep{{\fv{\it AGILE}}}
\shorttitle{Low frequency detections of \frb }
\shortauthors{Pilia et al.}
\graphicspath{{./}{figures/}}

\begin{document}

\title{The lowest frequency Fast Radio Bursts: Sardinia Radio Telescope detection of the periodic FRB~180916 at 328 MHz\footnote{Released on March, 27th, 2020}}

\correspondingauthor{Maura Pilia}
\email{maura.pilia@inaf.it}

\author[0000-0001-7397-8091]{M. Pilia}
\author[0000-0002-8265-4344]{M. Burgay}
\affiliation{INAF - Osservatorio Astronomico di Cagliari - via della Scienza 5 - I-09047 Selargius, Italy}
\author[0000-0001-5902-3731]{A. Possenti}
\affiliation{INAF - Osservatorio Astronomico di Cagliari - via della Scienza 5 - I-09047 Selargius, Italy}
\affiliation{Università di Cagliari - Dipartimento di Fisica - S.P. Monserrato-Sestu Km 0,700 - 09042 Monserrato, Italy}
\author[0000-0001-6762-2638]{A. Ridolfi}
\affiliation{INAF - Osservatorio Astronomico di Cagliari - via della Scienza 5 - I-09047 Selargius, Italy}
\affiliation{Max Planck Institute f\"ur Radioastronomie, Auf dem H\"ugel 69, D-53121 Bonn, Germany}
\author[0000-0002-8604-106X]{V. Gajjar}
\affiliation{Department of Astronomy,  University of California Berkeley, Berkeley CA 94720}
\author[0000-0002-5924-3141]{A. Corongiu}
\author[0000-0002-1806-2483]{D. Perrodin}
\affiliation{INAF - Osservatorio Astronomico di Cagliari - via della Scienza 5 - I-09047 Selargius, Italy}
\author[0000-0002-0916-7443]{G. Bernardi}
\affiliation{INAF-Istituto di Radio Astronomia, via Gobetti 101, 40129 Bologna, Italy}
\affiliation{Department of Physics and Electronics, Rhodes University, PO Box 94, Grahamstown, 6140, South Africa}
\affiliation{South African Radio Astronomy Observatory, Black River Park, 2 Fir Street, Observatory, Cape Town, 7925, South Africa}
\author[0000-0003-1807-6188]{G. Naldi}
\author[0000-0003-2172-1336]{G. Pupillo}
\affiliation{INAF-Istituto di Radio Astronomia, via Gobetti 101, 40129 Bologna, Italy}
\author[0000-0001-7915-996X]{F. Ambrosino}
\affiliation{INAF/IAPS, via del Fosso del Cavaliere 100, I-00133 Roma (RM), Italy}
\affiliation{Sapienza Universit\`a di Roma, Piazzale Aldo Moro 5, I-00185 Roma (RM), Italy}
\author{G. Bianchi}
\affiliation{INAF-Istituto di Radio Astronomia, via Gobetti 101, 40129 Bologna, Italy}
\author[0000-0002-8734-808X]{A. Burtovoi}
\affiliation{Centre of Studies and Activities for Space (CISAS) "G. Colombo", University of Padova, Via Venezia 15, 35131 Padova, Italy}
\affiliation{INAF - Osservatorio Astronomico di Padova - Vicolo dell'Osservatorio 5 - 35122 Padova, Italy}
\author[0000-0002-0752-3301]{P. Casella}
\affiliation{INAF/OAR, via Frascati 33, I-00078 Monte Porzio Catone (RM), Italy}
\author[0000-0001-8100-0579]{C. Casentini}
\affiliation{INAF/IAPS, via del Fosso del Cavaliere 100, I-00133 Roma (RM), Italy}
\affiliation{INFN Sezione di Roma 2, via della Ricerca Scientifica 1, I-00133 Roma (RM), Italy}
\author{M. Cecconi}
\affiliation{Fundaci\'on Galileo Galilei - INAF - Rambla J.A.Fern\'andez P. 7, 38712, S.C.Tenerife, Spain}
\author{C. Ferrigno}
\affiliation{ISDC, Department of Astronomy, University of Geneva, Chemin d’Écogia, 16 CH-1290 Versoix, Switzerland}
\author[0000-0002-7352-6818]{M. Fiori}
\affiliation{Department of Physics and Astronomy, University of Padova, Via F. Marzolo 8, 35131, Padova, Italy}
\author{K. C. Gendreau}
\affiliation{Astrophysics Science Division, NASA Goddard Space Flight Center, Greenbelt, MD 20771, USA}
\author{A. Ghedina}
\affiliation{Fundaci\'on Galileo Galilei - INAF - Rambla J.A.Fern\'andez P. 7, 38712, S.C.Tenerife, Spain}
\author[0000-0003-2007-3138]{G. Naletto}
\affiliation{Department of Physics and Astronomy, University of Padova, Via F. Marzolo 8, 35131, Padova, Italy}
\affiliation{INAF - Osservatorio Astronomico di Padova - Vicolo dell'Osservatorio 5 - 35122 Padova, Italy}
\author[0000-0001-8534-6788]{L. Nicastro}
\affiliation{INAF—Osservatorio di Astrofisica e Scienza dello Spazio di Bologna, Via Piero Gobetti 93/3, I-40129 Bologna, Ita}
\author[0000-0001-5578-8614]{P. Ochner}
\affiliation{Department of Physics and Astronomy, University of Padova, Via F. Marzolo 8, 35131, Padova, Italy}
\affiliation{INAF - Osservatorio Astronomico di Padova - Vicolo dell'Osservatorio 5 - 35122 Padova, Italy}
\author[0000-0002-8691-7666]{E. Palazzi}
\affiliation{INAF—Osservatorio di Astrofisica e Scienza dello Spazio di Bologna, Via Piero Gobetti 93/3, I-40129 Bologna, Ita}
\author[0000-0003-0543-3617]{F. Panessa}
\affiliation{INAF/IAPS, via del Fosso del Cavaliere 100, I-00133 Roma (RM), Italy}
\author{A. Papitto}
\affiliation{INAF/OAR, via Frascati 33, I-00078 Monte Porzio Catone (RM), Italy}
\author[0000-0001-6661-9779]{C. Pittori}
\affiliation{SSDC/ASI, via del Politecnico snc, I-00133 Roma (RM), Italy}
\affiliation{INAF/OAR, via Frascati 33, I-00078 Monte Porzio Catone (RM), Italy}
\author[0000-0003-2177-6388]{N. Rea}
\affiliation{Institute of Space Sciences (ICE, CSIC), Campus UAB, Carrer de Can Magrans s/n, 08193, Barcelona, Spain}
\affiliation{Institut d'Estudis Espacials de Catalunya (IEEC), Carrer Gran Capit\`a 2--4, 08034 Barcelona, Spain}
\author[0000-0003-3952-7291]{G. A. Rodriguez Castillo}
\affiliation{INAF/OAR, via Frascati 33, I-00078 Monte Porzio Catone (RM), Italy}
\author[0000-0001-6353-0808]{V. Savchenko}
\affiliation{ISDC, Department of Astronomy, University of Geneva, Chemin d’Écogia, 16 CH-1290 Versoix, Switzerland}
\author{G. Setti}
\affiliation{Dipartimento di Fisica e Astronomia, Universit\'{a} di Bologna, Via Gobetti 93/2, 40129 Bologna, Italy}
\affiliation{INAF-Istituto di Radio Astronomia, via Gobetti 101, 40129 Bologna, Italy}
\author[0000-0003-2893-1459]{M. Tavani}
\affiliation{INAF/IAPS, via del Fosso del Cavaliere 100, I-00133 Roma (RM), Italy}
\affiliation{Università degli Studi di Roma "Tor Vergata", via della Ricerca Scientifica 1, I-00133 Roma (RM), Italy}
\author[0000-0002-3180-6002]{A. Trois}
\affiliation{INAF - Osservatorio Astronomico di Cagliari - via della Scienza 5 - I-09047 Selargius, Italy}
\author[0000-0002-1530-0474]{M. Trudu}
\affiliation{INAF - Osservatorio Astronomico di Cagliari - via della Scienza 5 - I-09047 Selargius, Italy}
\affiliation{Università di Cagliari - Dipartimento di Fisica - S.P. Monserrato-Sestu Km 0,700 - 09042 Monserrato, Italy}
\author[0000-0002-9719-3157]{M. Turatto}
\affiliation{INAF - Osservatorio Astronomico di Padova - Vicolo dell'Osservatorio 5 - 35122 Padova, Italy}
\author{A. Ursi}
\affiliation{INAF/IAPS, via del Fosso del Cavaliere 100, I-00133 Roma (RM), Italy}
\author[0000-0003-3455-5082]{F. Verrecchia}
\affiliation{SSDC/ASI, via del Politecnico snc, I-00133 Roma (RM), Italy}
\affiliation{INAF/OAR, via Frascati 33, I-00078 Monte Porzio Catone (RM), Italy}
\author[0000-0002-6516-1329]{L. Zampieri}
\affiliation{INAF - Osservatorio Astronomico di Padova - Vicolo dell'Osservatorio 5 - 35122 Padova, Italy}





\begin{abstract}

We report on the lowest-frequency detection to date of three bursts from FRB~180916.J0158+65, observed at 328 MHz with the Sardinia Radio Telescope (SRT). The SRT observed the periodic repeater FRB~180916.J0158+65 for five days from Feb. 20, 2020 to Feb. 24, 2020 during a time interval of active radio bursting, and detected the three bursts during the first hour of observations; no more bursts were detected during the remaining $\sim 30$ hours. Simultaneous SRT observations at 1548 MHz did not detect any bursts. Burst fluences are in the range 13 to 37 Jy ms. No relevant scattering is observed for these bursts. 

We also present the results of the multi-wavelength campaign we performed on FRB~180916.J0158+65, during the $\sim 5$ days of the active window. Simultaneously with the SRT observations, observations with different time spans were performed with the Northern Cross at 408 MHz, with {\it XMM-Newton}, NICER, INTEGRAL, AGILE and with the TNG and two optical telescopes in Asiago, which are equipped with fast photometers.
{\it XMM-Newton} obtained data simultaneously with the three  bursts detected by the SRT, and determined a luminosity upper limit in the 0.3--10~keV energy range of $\sim10^{45}$~erg\,s$^{-1}$  for the burst emission. 
AGILE obtained data simultaneously with the first burst and determined a fluence upper limit in the MeV range for millisecond timescales of $ 10^{-8} \rm \, erg \, cm^{-2}$.

 Our results  show that absorption from the circum-burst medium does not significantly affect the emission from \frb.J0158+65, thus limiting the possible presence of a superluminous supernova  around the source, and indicate that a cutoff for the bursting mechanism, if present, must be at lower frequencies. Our multi-wavelength campaign sensibly constrains the broad-band emission from \frb.J0158+65, and provides the best limits so far for the electromagnetic response to the radio bursting of this remarkable source of fast radio bursts.

\end{abstract}



\section{Introduction} 
\label{sec:intro}
Fast Radio Bursts (FRBs) are fast, millisecond-duration, extremely bright ($\sim $Jy) bursts that have so far only been observed at radio wavelengths. Their extragalactic nature has been confirmed by the first Repeating Fast Radio Burst (RFRB), FRB\,121102, located at a redshift of $z = 0.193$ \citep{spitler14,spitler16,tendulkar17,chatterjee17,marcote17}.  However, in the last year, observations performed in the frequency band between 400 and 800 MHz by the Canadian Hydrogen Intensity Mapping Experiment (CHIME, \citealt{chimefrb18}), led to a significant growth in the known population of ``Repeaters'' (alternate name for RFRBs) \citep{chime_rep19,chime8frb,chime9rep}. 

Among the 20 repeating FRB sources published until now,  FRB~180916.J0158+65 (hereafter FRB~180916) \citep{chime8frb} was discovered by CHIME through the detection of 10 bursts, with a flux density ranging from $\sim0.4$ to $\sim4$~Jy. A subsequent targeted VLBI campaign, favored by the active nature of the source and its low extra-galactic dispersion measure of ${\rm DM}\sim 349$~pc~cm$^{-3}$, led to the identification of the host galaxy at a redshift $z = 0.0337$ \citep{marcote20}. The localization of \frb, the second ever for a RFRB, immediately showed a dichotomy with the case of FRB\,121102: indeed, \frb\ was found in a star-forming region within a nearby massive spiral galaxy, at odds with FRB\,121102, which is hosted in a dwarf galaxy \citep{chatterjee17,marcote17}. The subsequent continuous monitoring of \frb\, by CHIME led to the first identification of a periodicity in the active phases of a RFRB (\citealt{chime_period_20}, hereafter CF20) (now possibly also followed by the detection of a periodicity from the original repeater, \citealt{rajwade20}). In particular, \frb\ displays a periodicity of $16.3$~days in its phases of activity, with an active window phase concentrated within $\pm2.6$~days around the midpoint of the window. Although it is still rather uncertain, the radio burst rate during the active window in the CHIME frequency band is $\sim 1.0\pm0.5$ per hour. The reported periodicity  seems far too long to be ascribed to a neutron star's rotational frequency,  unless FRB progenitors are older, ultra-long period magnetars \citep{beniamini20}. This triggered a wealth of alternative hypotheses, such as orbital effects \citep{lyutikov20,iokazhang20} or various kinds of precessional effects \citep{yang_zou20,levin+20, zanazzi_lai20,gu+20}, including the precession of a jet produced by intermediate black hole accretion \citep{katz20}, or other secular semi-periodic cyclic phenomena, e.g. the source traveling across an asteroid belt \citep{dai_zhong20}. Most of these models revisited concepts developed for the modeling of the original RFRB (for a review see \citealt{platts19}), tuned to a $\sim16$ days periodicity.

From an observational point of view, the availability of predictable ``windows for radio observations", in combination with the rare proximity of the source, makes \frb\ the best target for additional studies in the radio band, as well as for multi-wavelength follow-ups. We have exploited one of these windows (the one centered on 21 February 2020) in order to search, with the Sardinia Radio Telescope (SRT; \citealt{bolli+15,prandoni17}), for the signature of bursts at frequencies below 400 MHz, where no FRB nor RFRB have so far been detected (\citealt{rajwade20a,houben19}). In parallel, we set up observations at higher radio frequencies, as well as in the optical, X-ray and $\gamma$-ray bands.

In \S2 we describe the multi-wavelength campaign while in \S3 we present the results, focused on the detection of three bursts at 328 MHz with the SRT, and on the upper limits (ULs) from the simultaneous observations at other wavelengths; in \S4 we discuss the properties of the bursts and some implications resulting from this initial campaign. 

\section{Observations} \label{sec:Obs}

We carried out a multi-wavelength campaign to look at \frb\, during its active cycle starting on 2020-02-19, centered on 2020-02-21 at 16:12 UT, and ending on 2020-02-24. 
The campaign was set up in order to maximize the overlap of the simultaneous observations between the radio observatories and the multi-wavelength instruments. Details of the observational campaign can be found in Fig. \ref{fig:obscampaign}.

\subsection{Radio observations}
\subsubsection{The Sardinia Radio Telescope} \label{subsec:radio}

The 64-m SRT  observed \frb\ for a total of 30 h over a time span of five days. 
Observations were performed using the L/P dual-band coaxial receiver \citep{valente+10}, with the two observing bands centered at 1548 MHz  (L-band) and 328 MHz (P-band), respectively. 
Observations at L-band were performed with the ATNF Digital Filterbank Mark III backend (DFB\footnote{\tt{www.jb.man.ac.uk/pulsar/observing/DFB.pdf}}), with a bandwidth of 500 MHz, 1-MHz-wide channels and a sampling time $t_{\rm s} = 125\,\mu$s. Observations in the P band were performed in baseband mode using the ROACH1 backend \citep{bassa+16}, which acquired data over a 64-MHz bandwidth. 

The modified radiometer formula for pulsars \citep{lk04} applied to the SRT observations at L and P bands results into a minimum detectable flux density of 600 mJy and 2.2 Jy for a 1\,ms burst, in the two bands, respectively. For P band we consider a $6 \sigma$ limit.
Given the presence of radio frequency interference (RFI) in the L band, the actual useful bandwidth was reduced to 350 MHz. Wide-band RFI, which in some cases saturated the backend, was also present; to take its effects into account, we set a threshold limit of $10\sigma$ for L-band searches.
The telescope gain in the two bands is, $G_L = 0.55$~K\,Jy$^{-1}$ and $G_P = 0.52$~K\,Jy$^{-1}$, respectively. The system temperatures, accounting for the antenna temperature and the sky temperature, as extrapolated from 408\,MHz all-sky map \citep{reich88}, are 30 K and 60 K, respectively. 

\begin{figure}
    \centering
    \includegraphics[scale=0.25]{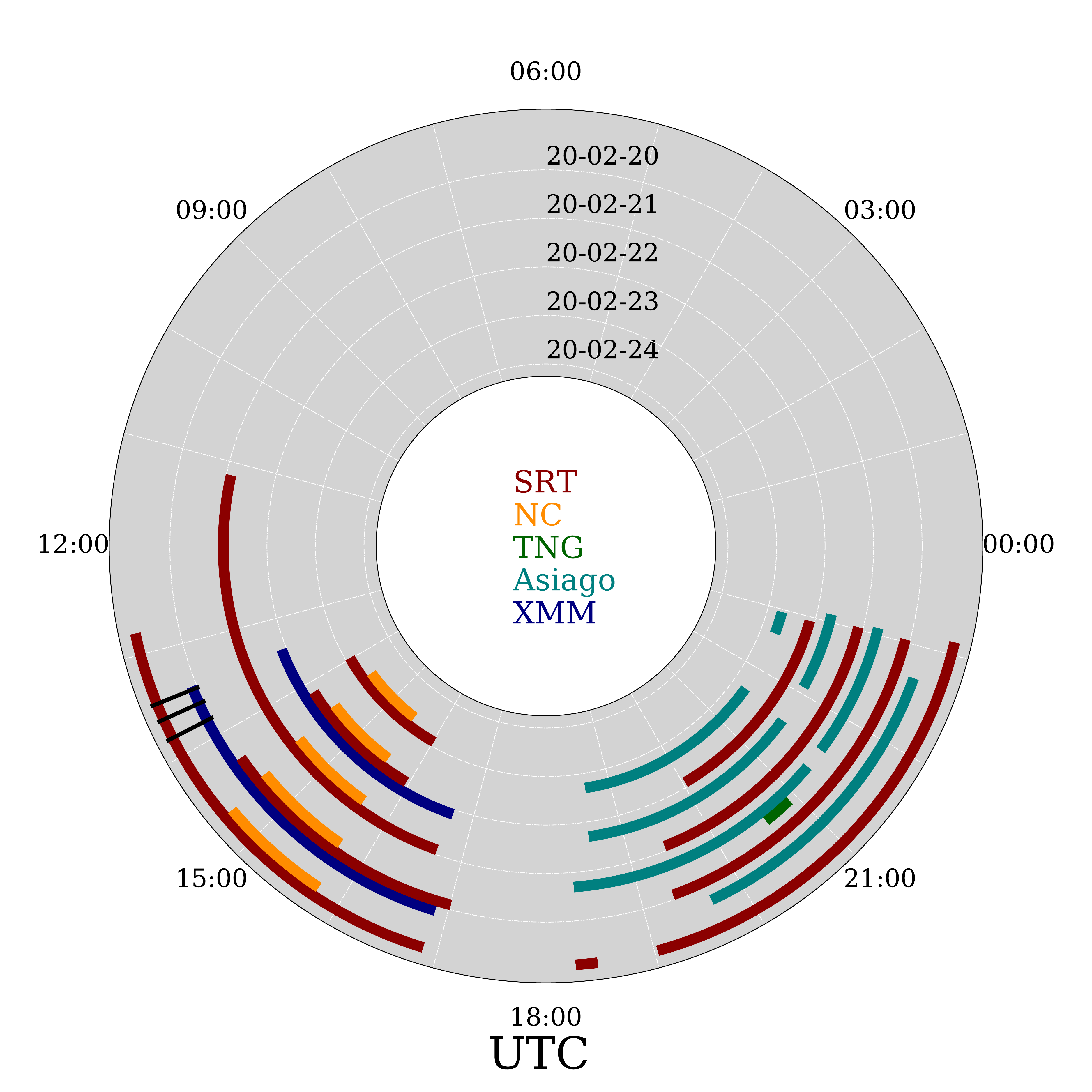}
    \caption{Observational multi-wavelength campaign around the SRT observations (red). Concentric circles represent different days. NICER, Integral and AGILE  were active during most of the SRT observing windows, except during epochs of visibility limitations (see text for details). The three black radial segments between UTC 13:00 and 14:00 represent the three bursts.}
    \label{fig:obscampaign}
\end{figure}

\subsubsection{The Northern Cross radio telescope} \label{subsec:radiocroce}

Observations with the Northern Cross (NC) radio telescope were carried out for seven days starting on 2020-02-19 and ending on 2020-02-25, for a total observing time of about 7.5~h (1.1 h per day).
The system used to observe FRBs with the NC is described in \citet{locatelli20} and will be briefly summarized here. Six cylinders of the North-South arm were used for our observations. Each cylinder is composed of four receivers. Signals from the 24 receivers were calibrated by observing Cas~A in interferometric mode, 
and then  combined together in a single digital beam that followed the source as it transited across the field of view (FoV). Beam-formed voltages were stored to disk with a 138.24~$\mu$s time resolution over a 16~MHz bandwidth centered at 408~MHz, with a 12.2~kHz frequency resolution. The increased frequency resolution, compared to the 781~kHz used in \citet{locatelli20}, was achieved by Fourier transforming each frequency channel in a time window of 64 points and led to negligible intrachannel smearing. 

\subsection{X-ray/Gamma-ray observations} \label{subsec:datax}

\subsubsection{XMM-Newton} \label{subsec:xmm}

We obtained two {\it XMM-Newton} observations allocated as Director Discretionary Time (OBs 0854590701 and 0854590801). 
The observations were performed on 2020-02-20 (from 13:27 to 16:52) and on 2020-02-22 (from 13:26 to 16:43). 
These on-source UT times refer to the EPIC-pn instrument, which observed the target in Full Frame and thin filter, with a time resolution of 73.37 ms. The on-source times for the other instruments (MOS cameras in Full Mode, thin filter; RGS in Default Spectroscopy Mode; OM in Fast Mode, U filter) vary only slightly. Data reduction was carried out with the SAS Data Analysis software version 17.0.0.

\subsubsection{NICER} \label{subsec:nicer}

NICER observed \frb\ with several short observations over 6 days, to cover the predicted activity period of the target, obtained as Director Discretionary Time. A total of more than 113 ks were collected on source, with a sub-$\mu$s time resolution in the 0.2-12 keV energy range. 

\subsubsection{INTEGRAL}

INTEGRAL observed \frb\ via a Target of Opportunity in search for a possible steady or impulsive hard X-ray/soft $\gamma$-ray counterpart to its radio emission. The INTEGRAL observations were interrupted by two spacecraft perigee passages, and were carried out from 2020-02-20 14:53:58 to 2020-02-21 05:12 UT, from 2020-02-21 15:38 to 2020-02-23 20:42 UT, and from 2020-02-24 06:39 to 2020-02-24 17:20:31 UT, for a total on-source time of 261 ks (3 days). 

All INTEGRAL data were processed using the standard INTEGRAL Offline Scientific Analysis (OSA) software, version 11.0.

\subsubsection{\agilep} \label{subsec:agile}

\agile observed FRB 180916 with two detectors:
the $\gamma$-ray imaging detector (GRID), which is sensitive in the range 30 MeV -- 30 GeV with a 2.5 sr FoV, and the Mini-Calorimeter (MCAL), which is sensitive in the 0.4--100 MeV band with $4\pi$ non-imaging acceptance \citep[][]{Tavani2009,Tavani2019}.
\agile is currently operating in spinning mode, with the instrument axis rotating every $\sim 7$ minutes around the satellite-Sun direction. For each satellite revolution, a large fraction of the sky ($\sim 40-60$\%) is exposed, depending on the Earth occultation pattern and trigger disabling over the South Atlantic anomaly (SAA, about $10$\% of the 95-min orbit). Over timescales of hours,
  $\sim$80\% of the entire sky can be exposed by the GRID $\gamma$-ray imager and by the MCAL.

\subsection{Optical observations} \label{subsec:optical}
\subsubsection{SiFAP2/TNG}
\frb\ was observed in the optical band with 
SiFAP2 \citep{ghedina18}
at the INAF's 3.58-m Telescopio Nazionale Galileo (TNG).
Based on the Silicon Photo Multiplier (SiPM) technology, SiFAP2 
is composed of two Multi Pixel Photon Counter (MPPC) sensors 
working in the optical band ranging from 320 to 900~nm \citep{meddi12,ambrosino16,ambrosino17}. 
Each sensor 
has a time tagging capability
of 8~ns
and can integrate the number of incoming photons in adjustable time windows ranging from 100~ms down to 1~ms. A commercial Global Positioning System (GPS) unit provides the absolute time 
with an accuracy that is better than 60~$\mu$s \citep{papitto19} on the UTC.
At the TNG focal plane, the FoV of each sensor is about 7 $\times$ 7~arcsec$^2$, ensuring that the sources are completely collected even in bad seeing conditions ($\sim$3~arcsec). The two MPPC detectors acquired simultaneously the target (FRB 180916) and the nearby sky background located at an angular distance of 4~arcmin away from the target itself.  
The SiFAP2 observing run was carried out from 2020-02-21 20:35:02 to 2020-02-21 20:54:31 UT, for a total exposure time of roughly 1.2~ks. The acquisition was stopped because of bad weather conditions, ensuring only a short window (less than 20 min) of simultaneous observation with Aqueye+ (see \S2.6.2) and the SRT. 

\subsubsection{Aqueye+ and IFI+Iqueye at Asiago}
\frb\ was also observed with Aqueye+, which is mounted at the Copernicus telescope, and IFI+Iqueye, which is mounted at the Galileo telescope in Asiago, Italy. Aqueye+ and Iqueye\footnote{https://web.oapd.inaf.it/zampieri/aqueye-iqueye/} are fast photon counters with a field of view of 6--12 arcsec and the capability of time tagging the detected photons with sub-ns time accuracy \citep{2009JMOp...56..261B,2009A&A...508..531N,2013SPIE.8875E..0DN,2015SPIE.9504E..0CZ}. Iqueye is fiber-fed through a dedicated instrument (Iqueye Fiber Interface; \citealt{2019CoSka..49...85Z}). We also performed simultaneous observations of the field in the sloan $i$ band with a conventional CCD camera mounted on the 67/92 Schmidt telescope. Several (unfiltered) acquisitions were performed with Aqueye+ and IFI+Iqueye between 2020-02-20 and 23 (see Fig. \ref{fig:obscampaign}),
for a total
on-source time of $\sim$11.5 h for Aqueye+ and $\sim$13.5 h for IFI+Iqueye. 
The sky background was simultaneously and continuously monitored with the on-sky detector of Aqueye+ (approximately 10 arcmin away from the target and with a FoV comparable to that of the on-source detectors). The average count rate measured with the on-source detectors was 2900--4300 count/s for Aqueye+ and 1800--3200 count/s for IFI+Iqueye. The data reduction was performed with dedicated software. The whole acquisition and reduction chain ensure an absolute accuracy of $\sim$0.5 ns relative to UTC \citep{2009A&A...508..531N}.

\section{Data Analysis and Results} \label{sec:Results}

\subsection{Radio} \label{subsec:resultadio}

\subsubsection{SRT} 

L-band data were recorded at 2-bits per sample by the DFB as {\tt psrfits} \citep{hotan04} files; these were later converted to 8-bit filterbank files using {\tt SIGPROC} \citep{lorimer11}. P-band data were acquired as {\tt dada} baseband files and subsequently converted to 8-bit filterbank format  using {\tt digifil} \citep{dspsr}.
 The 64-MHz bandwidth was divided into 256, 250-kHz-wide channels and the resulting $4~\mu$s time resolution was then averaged down to $128~\mu$s; the data were coherently dedispersed at the nominal ${\rm DM} = 348.82$~pc\,cm$^{-3}$ (CF20). 
The P-band channelized filterbank files were processed through the Python-based pipeline named {\tt SPANDAK}\footnote{\url{https://github.com/gajjarv/PulsarSearch}}, which is similar to the one used in \cite{Gaj18}. Data were first processed through {\tt rfifind} from the {\tt PRESTO} package\footnote{\url{https://www.cv.nrao.edu/~sransom/presto/}} for high-level RFI purging.
The pipeline uses
{\tt Heimdall} \citep{bbb+12} as the main kernel to quickly search across a DM range from 300 to 400 pc~cm$^{-3}$. Since Heimdall is unaware of the fact that the data is coherently de-dispersed at the nominal DM of the source, we used a threshold of 0.01\% for the maximum sensitivity loss for each given DM, so that the DM step would be 0.03 pc cm$^{-3}$, corresponding to a maximum DM smearing of 0.5 ms across the observing band. For the same reason, we used the option {\tt -no\_scrunching} to avoid time rebinning at the FRB's DM.
The de-dispersed time-series were searched for pulses using a matched-filtering technique with a maximum window size of 32.8 ms. Each candidate found by {\tt Heimdall} was scrutinized against all other candidates for each given observation to validate and identify only the genuine ones. 
 The pipeline produced around 7000 candidates at different DMs. All candidates with DM between 340 and 360 pc cm$^{-3}$ were visually inspected; 
we identified three clear bursts from the observations at 328 MHz on 2020-02-20, which will be discussed below.
As a cross-check, we also analyzed the P-band data using {\tt PRESTO} over 121 DM values covering the range 345.82 -- 351.82 pc cm$^{-3}$. A slightly different approach was used for RFI excision, where only the frequency channels affected by strong RFI were removed through the {\tt -ignorechan} option of the {\tt PRESTO}'s {\tt prepsubband} routine. The option {\tt -noclip} was also used during de-dispersion to avoid strong bursts being flagged as RFI. The python code {\tt single\_pulse\_search.py} was used with a signal-to-noise (S/N) threshold of 6, a maximum width of 38.4 ms, and the option {\tt -b} to avoid checking for bad blocks (potentially saving strong pulses from being discarded). The only good single pulses found were the same three as the ones detected by the {\tt SPANDAK} pipeline. The remaining candidates were all monochromatic bursts of RFI.

L-band observations were analyzed using the same pipeline as the P-band observations and, given the large amount of RFI, using the single pulse search from {\tt PRESTO} on a small DM range around the nominal value. In particular, the data were first cleaned from RFI using {\tt rfifind}, then de-dispersed with {\tt prepsubband} using 12 DM values from 346.32 to 351.82. The resulting time series were analyzed with {\tt single\_pulse\_search.py} using a S/N  threshold of 10 and a maximum width of 37.5 ms. All of the resulting candidates were visually inspected and recognized as RFI.  

In summary, the SRT 
detected three radio bursts from FRB 180916 in the P band
on 2020-02-20. 
It is interesting to note that this is the 
first firm detection of a FRB below 400 MHz. Detection of nine bursts at 111 MHz has been reported by \cite{fedrod19}. However, the observational setup of their system (a coarse frequency resolution of 78~kHz over a tiny observing bandwidth of 2.5~MHz, and a sampling time not faster than 12.5~ms) required the use of a template matching approach in order to see the bursts. Although the authors do their best to support the validity of this methodology, its use is very limited so far in the context of FRB searches, and the statistics of the false-positive is not completely assessed. Moreover, the claimed detection is very hard to reconcile with the stringent limits imposed by all other non-detections at similar frequencies, derived by using well consolidated procedures \citep{Coenen14, Tingay15, Karastergiou15, sokolowski,chawla20}.
Properties of these bursts are summarized in Table~\ref{tab:tab1}. No simultaneous bursts (taking into account the DM-delay between the two bands) were detected blindly in the L-band data, down to a limiting sensitivity of $600 \times (W / {\rm ms})^{-0.5}$ mJy, where $W$ is the pulse width in ms. By analyzing the relevant data segments, the limit can be moved down to $360 \times (W /{\rm ms})^{-0.5}$ mJy, with a $6 \sigma$ threshold. 

\begin{deluxetable*}{ccccccc}
\tablecaption{Properties of the three 
radio bursts from FRB 180916 detected by the SRT. Burst arrival times are barycentric at infinite frequency, i.e. after correcting for the DM delay at 328 MHz.
 All bursts were detected on Feb. 20, 2020.}
\label{tab:tab1}
\tablehead{
\colhead{Time} & \colhead{Time} & \colhead{Width}  & \colhead{S/N} & \colhead{Peak flux}  & \colhead{Fluence}  & \colhead{DM}  \\
\colhead{(UT)} & \colhead{(MJD)} & \colhead{(ms)} & \colhead{(-)}  & \colhead{(Jy)} & \colhead{(Jy ms)} & \colhead{(pc cm$^{-3}$)}\\
}
\startdata
13:28:25.983(8) & 58899.56141184 & 13(4) & 31.7 & 2.8(9) & 37(16) & 349.8(1)\\
13:37:39.437(7) & 58899.56781756 & 9(4) & 13.6 & 1.5(7) & 13(8) &349.4(1) \\
13:48:53.20(1) & 58899.57561573 & 14(4) & 16.0 & 1.4(4) & 19(8) &350.1(1)\\
\enddata
\end{deluxetable*}

\begin{figure*}
    \includegraphics[width=1\textwidth]{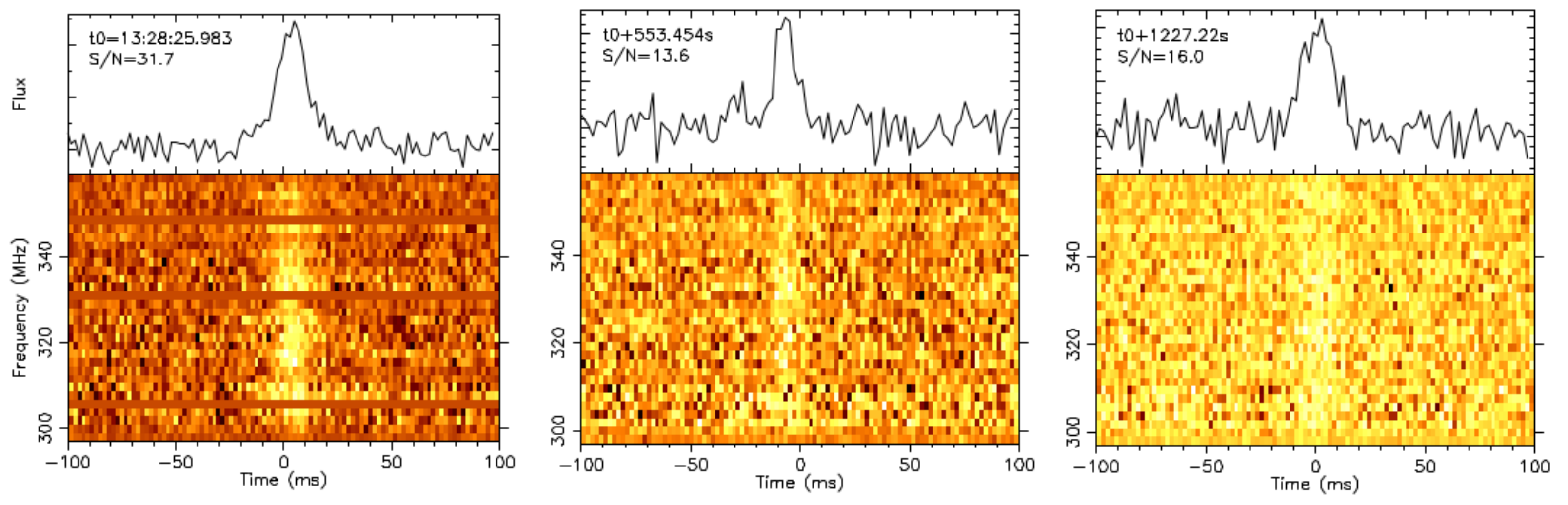}
    \caption{Profiles of the three bursts observed by the SRT. The upper panels represent the pulse profile in arbitrary flux units; the lower panels show the dynamic spectra of the bursts, here scrunched into 32 frequency channels. Horizontal blank lines represent channels zapped due to RFI.}
    \label{fig:bursts}
\end{figure*}

\subsubsection{DM optimization and Burst structures}
FRBs are known to show complex burst structures with multiple burst components. Many of the repeating and non-repeating FRBs have been shown to exhibit a drifting emission pattern where emission gradually moves from higher to lower frequencies across the leading component to consecutive trailing components. \cite{Gaj18} showed that these burst structures can superimpose for different trial DMs, which could lead to incorrect DM estimations. As shown in Fig. \ref{fig:bursts}, all of our detected bursts show a single component. However, it is possible that our observations were not sensitive enough to resolve these underlying structures. One of the ways to reveal such structures is to estimate ``structure-maximizing DM'' -- DM where all sub-burst structures spanning different frequencies arrive at the same time. This structure-maximizing DM can then be compared with the S/N-maximizing DM (DM where S/N of integrated burst profile peaks). If there are underlying structures, they are likely to superimpose to give higher S/N-maximizing DM compared to the structure-maximizing DM (see Figure 1 in \citealt{Gaj18}). 

\begin{figure*}[t]
    \centering
    \includegraphics[scale=0.7]{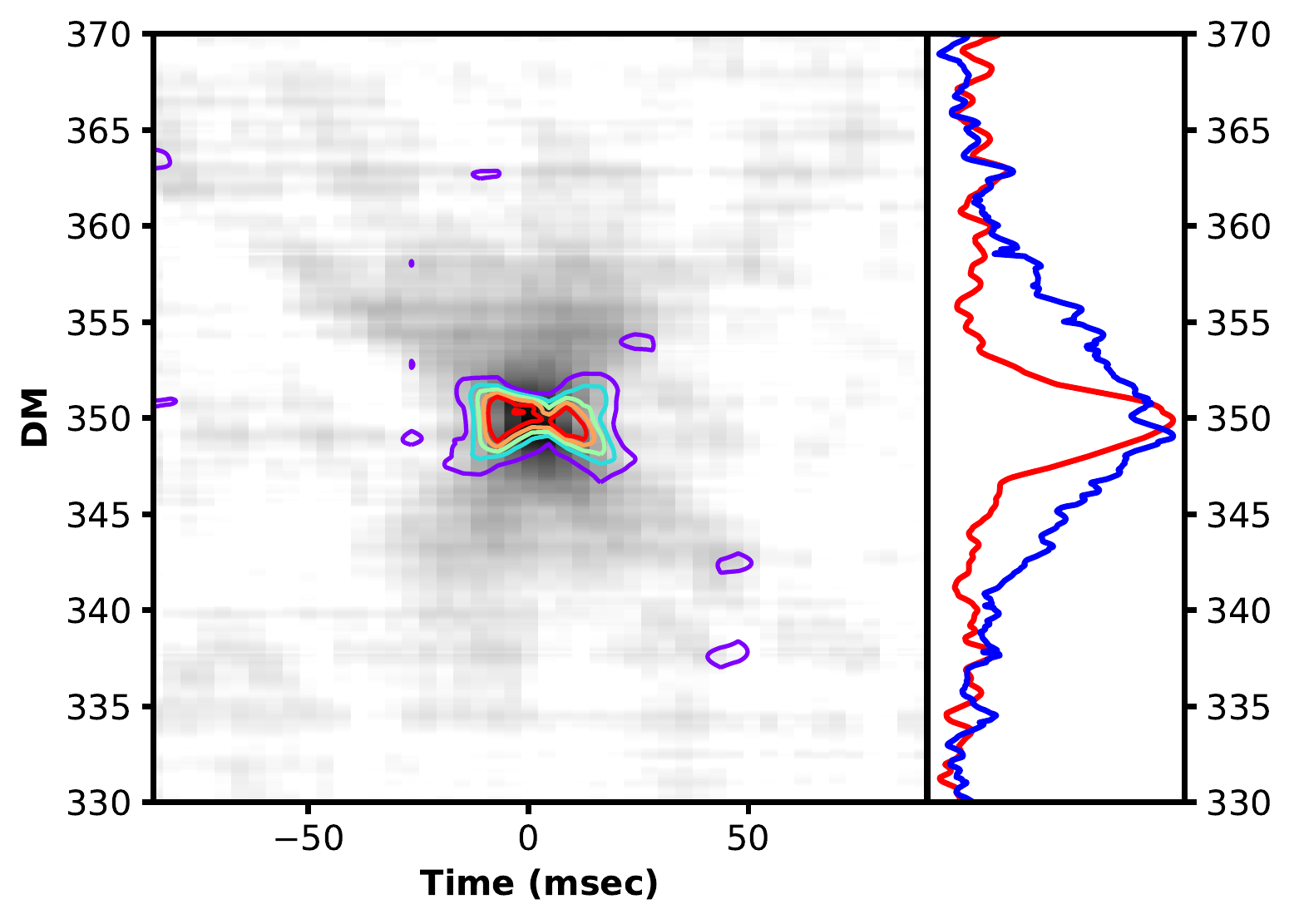}
    \caption{Comparison of S/N-maximizing DMs with the structure-maximizing DMs for burst-1 detected across 300 to 360 MHz at the SRT. The left panel shows the S/N in the gray-scale intensities as a function of DM-vs-time in the background. The standard `butterfly' pattern is clearly visible with a peak in the middle. The overlaid contours (with levels at 0.35, 0.5, 0.6, 0.7, and 0.8 times the peak) are from the structure-maximizing parameter intensities. The right panel shows the normalized average intensities from a -20 to +20 msec region around the peak; blue and red lines represent S/N and structure parameter as a function of DM, respectively. The original data, with a temporal resolution of 244 $\mu$s and a spectral resolution of 7.8 khz, were first averaged to obtain temporal and spectral resolution of 4\,ms and 1 MHz, respectively. We also used a uniform 2D filter of size 0.1 pc-cm$^{-3}$ $\times$ 11 ms to smooth the intensities. A Gaussian function was fitted for both average intensities in the right hand panel. We found a S/N-maximizing DM of 350.1(0.3) pc-cm$^{-3}$ and a structure-maximizing DM of 349.8(0.1) pc-cm$^{-3}$. Here, uncertainties were obtained from standard deviations of DM and S/N intensities from the off-pulse and after propagating them during Gaussian fitting.}
    \label{fig:dm-snr}
\end{figure*}

Two different techniques have been proposed to estimate the structure-maximizing DM. \cite{chime_discovery} suggested a coherent summation of Fourier transformed spectra taken across channels for the burst, while \cite{Gaj18} suggested maximizing the forward time-derivative of the burst profile. For weaker burst pulses, \cite{hss+19} suggested a second-order forward derivative, while \cite{jcf+19} suggested a fourth-order forward derivative. Here, we carried out a comparison between S/N-maximizing DMs and structure-maximizing DMs across a range of trial DMs for our brightest detected burst (i.e. burst-1). To estimate the structure-maximizing DM, we found that the second-order derivative was able to provide a single prominent peak. Fig. \ref{fig:dm-snr} shows the comparison of S/N-maximizing DMs with structure-maximizing DMs for burst-1. We did not find any significant difference between these two DMs within the measurement uncertainties. Thus, we confirm the absence of any underlying sub-burst structure for our detected bursts. This could also be due to our limited observing bandwidth. For example, burst number 2 (from day 181019) from \frb\  appears to show two components where the frequency span of each sub-burst structure is around 100 MHz \citet{chime}, which is larger than the bandwidth of our observations. 

\subsubsection{Northern Cross} 

Single-beam, channelized observations were analyzed following a pipeline similar to the SRT case and based on the {\tt Heimdall} code. Observations, which were not simultaneous with the SRT detected bursts, achieved a $3 \, (W/{\rm ms})^{-0.5}$~Jy $6\sigma$ rms sensitivity at 408~MHz. 
No burst was detected in the 7.5~h of campaign.

\subsection{X-ray and Gamma-ray observations} \label{subsec:resultx}
 
 \paragraph{\bf XMM-Newton} 
 
We focused on the data obtained by the EPIC-pn, since it has the largest effective area among the available instruments, and offered the best time resolution ($73.3$~ms compared to $2.7$~s for the EPIC-MOS cameras). We only considered the first hour of the first observations, during which the SRT detected the three bursts reported here; it was the only time in which the observations were not affected by a high flaring particle background.
We extracted good events, singles and doubles ({\texttt{PATTERN}} $\le 4$) with an energy range between 0.3 and 10 keV. We extracted a region of 25 arcseconds around the position of \frb, while different background regions were extracted on source-free parts of the same CCD, and then back-scaled to the dimension of the on-source extraction. At the position of the source, we have counted $30.0\pm5.5$ photons in 3.2~ks, which are compatible with the average of $31.8\pm0.9$ background photons.
The corresponding UL on the background subtracted average count rate, evaluated for a $3\sigma$ confidence level, is  $5.8\times10^{-3}$~s$^{-1}$ \citep[see][]{gehrels1986}. We assumed a spectral model typical of magnetar steady emission \citep[see, e.g.][]{rea2008}, composed of a black body (with $kT=0.5$~keV) and a power-law component (with $\Gamma=3$), and an absorption column of $4.3\times10^{21}$~cm$^{-1}$ evaluated from the Galactic H I Column Density maps in the source direction with the Heasoft {\texttt NH} tool. Under these assumptions, we estimated an upper limit on the unabsorbed {\it persistent} flux of $6\times10^{-14}$~erg~cm$^{-2}$~s$^{-1}$ (0.3--10~keV). At a luminosity distance of 149~Mpc, this corresponds to $\sim1.5\times10^{41}$~erg~s$^{-1}$.

A search for any impulsive or nearly-impulsive excess at or close to the times of the three bursts yielded no significant detection. The closest in time X-ray photon from the source position lagged the third observed radio burst by $\sim8$~s. We evaluated ULs of 90.3 $s^{-1}$ and 6.6~s$^{-1}$ on the count rate in each of the 73ms-bins coincident with radio detections, and within an interval of 1~$s$, respectively. Assuming a spectrum typical of SGR bursts \citep[see, e.g.][]{israel2008}, composed of two black body components with temperatures $kT_1=3$ and $kT_2=7$~keV, respectively, and the absorption column reported above, an upper limit on the unabsorbed {\it burst-like} (0.3--10~keV) flux of $2.2\times10^{-9}$~erg~cm$^{-2}$~s$^{-1}$ and of $1.6\times10^{-10}$~erg~cm$^{-2}$~s$^{-1}$ for the first and second radio detections. These correspond to luminosity limits to the X-ray counterparts of the two bursts of $\sim5.5\times10^{45}$~erg~s$^{-1}$ and $\sim4\times10^{44}$~erg~s$^{-1}$.

We note that by loosening the filtering criteria, namely including the lowest energies, a cluster of photons is detected at a position that is marginally consistent with the target, at UT time 2020-02-20 13:31:56 (barycentered to the Solar System using DE-405 ephemerides), which would correspond to either a few minutes before the first SRT burst, or a few minutes after the second SRT burst. However, a careful inspection of the data reveal that the detected photons form a clear track on the detector. We conclude that they are caused by the interaction of an energetic particle with the detector.

\paragraph{\bf NICER} NICER was not pointing at the target at the times SRT detected the three bursts reported here, because of viewing limitations. A search for impulsive events yielded no significant detection through the whole dataset. Further deeper searches are ongoing and the results will be presented in future publications.


\paragraph{\bf INTEGRAL} All three radio bursts reported in this paper occurred between $\sim 1$ and $\sim 1.5$~h before the start of the INTEGRAL pointing observation, and no deep ULs can be obtained close to the time of the discovered events. Using INTEGRAL all-sky detectors, we derive a 3-sigma UL on a 75-2000 keV fluence of any burst shorter than 1-s (50-ms) of $1.8 \times 10^{-7}$~erg~cm$^{-2}$ ($4 \times 10^{-8}$~erg~cm$^{-2}$) anywhere within $\pm 300$~seconds from the radio bursts. 
In addition, we searched for any short magnetar-like bursts in the entire INTEGRAL/ISGRI observation, and did not detect any, setting a 3-sigma UL on 28--80~keV fluence in 1~s at the level of $2.3\times10^{-8}$~erg~cm$^{-2}$, and on fluence in less than 100 ms at the level of $6.1 \times 10^{-9}$~erg~cm$^{-2}$. 

Integrating over the entire INTEGRAL exposure time, we do not detect any steady emitting source at the position of \frb. We derive a 3-sigma UL on the average flux of $3.1 \times 10^{-11}$~erg~cm$^{-2}$~s$^{-1}$ in the 28-80 keV energy range (with IBIS/ISGRI), and $2.3 \times 10^{-11}$~erg~cm$^{-2}$~s$^{-1}$ in the 3--10 keV range (with JEM-X).

 \paragraph{\bf AGILE} 
The \agilep/MCAL on-board data acquisition is based on a trigger logic acting on different energy ranges and timescales (ranging from $\sim300\, \mu$s to $\sim8$~s).
A detailed discussion about MCAL triggering and UL capabilities in the context of FRB studies  is reported in \citep[][]{Casentini2020,Ursi2019}. 

We searched for MCAL triggered events at or near the
radio bursts.
Our search was within $\pm 100$ seconds from the arrival times of Table \ref{tab:tab1}. 
MCAL collected useful data only for burst-1.
For the other two events, the AGILE satellite was in the SAA region, and no MCAL data could be obtained.
No significant event was detected in temporal coincidence with burst-1. The MeV fluence UL obtained for a millisecond timescale trigger is $F'_{MeV,UL} = 10^{-8} \rm \, erg \, cm^{-2}$. Table \ref{tab:tab3} shows the MCAL fluence ULs at different trigger timescales.  The value at 1 s corresponds to an isotropic MeV luminosity of $L_{MeV,UL} = 3.4\,\times\,10^{46} \rm \, erg \,s^{-1}$.

\begin{table*}[ht!]
  \begin{center}
    \caption{Average \agilep/{{MCAL}} fluence ULs in  {$\rm erg \, cm^{-2}$}}
    \begin{tabular}{|c|c|c|c|c|c|c|}
     \hline
    sub-ms & 1 ms & 16 ms & 64 ms & 256 ms & 1024 ms & 8192 ms\\
     \hline

     $1.13 \times 10^{-8}$ & $1.29 \times 10^{-8}$ & $3.72 \times 10^{-8}$ & $4.97 \times 10^{-8}$ & $7.95 \times 10^{-8}$ & $1.59 \times 10^{-7}$ & $4.49 \times 10^{-7}$\\

     \hline
    \end{tabular}
    \label{tab:tab3}
 \end{center}
\end{table*}

The analysis of $\gamma$-ray GRID data is based on the standard AGILE-GRID multi-source likelihood analysis \citep[][]{Bulgarelli2012} that takes into account nearby known $\gamma$-ray sources and the diffuse Galactic emission at the \frbp location ($l = 129.7$, $b = 3.7$).
 For this campaign, we obtained $\gamma$-ray flux ULs for different integrations: (1) a 100 s integration centered at the burst time of event-1; and (2) a 5-day integration covering a complete activity cycle of \frb (19-24 Feb., 2020); (3) a 30-day interval that also includes the previous cycle (4-8 Feb., 2020) during which AGILE was observing simultaneously with Swift \citep{tavani20}. The corresponding UL values are: $F_{\gamma,UL} = 2.7\, \times 10^{-7}\,\rm erg \, cm^{-2} \, s^{-1}$ for the 100 s integration above 50 MeV;  $F_{\gamma,UL} = 1.4\,\times\,10^{-10} \rm \, erg \, cm^{-2} \, s^{-1}$ for the 5-day integration above 100 MeV, and $F_{\gamma,UL} = 2.7\,\times10^{-11} \rm \, erg \, cm^{-2} \, s^{-1}$ for the 30-day integration above 100 MeV. The latter value corresponds to an isotropic long-term averaged $\gamma$-ray luminosity $L_{\gamma,ave,UL} \sim 7.2 \, \times\, 10^{43} \, \rm erg \, s^{-1}$.

\subsection{Optical} \label{subsec:resultoptical}
The optical telescopes did not observe simultaneously with the detected radio bursts, which happened during the day time.

 \paragraph{\bf SiFAP2} 
To derive an upper limit for the other slots of observation in this campaign, we considered the ratio between the count rate observed from the source position and from the sky background. This smoothed out the effect of quickly varying atmospheric conditions that characterized the sky during the observations. We did not detect any impulsive signal above a sensitivity threshold of 38.6 counts per 1 ms-long bin (3$\sigma$ confidence level after taking into account for the number of trials). This corresponds to a magnitude of $V\sim 15.3-15.5$.

\paragraph{\bf Aqueye+ and IFI+Iqueye}

We performed a systematic search for any significant increase in the count rate on 1-ms binned light curves of the optical observations (not background subtracted). We did not detect any simultaneous on-source radio-optical burst. Considering the average rate and the number of trials, the limiting sensitivity for the detection of a 1 ms pulse at the $\sim 3\sigma$ confidence level in an observation of 1 hour duration is 19--23 counts/bin for Aqueye+ and 15--20 counts/bin for Iqueye, corresponding to an instantaneous (1 ms) magnitude $V=13.4-13.7$ (fluence 0.012--0.016 Jy ms) for Aqueye+ a $V=11.7-12$ (fluence 0.060-0.079 Jy ms) for Iqueye.

Standard data reduction was applied to the simultaneous Schmidt images. No event was detected in any of the images down to a limiting magnitude $i=20$ (fluence 7.9 Jy ms for a 5-min exposure).

\section{Discussion} \label{sec:Discussion}

\subsection{Temporal distribution of the bursts} \label{subsec:radioburstsrate}
SRT observed for a total of 30.3 hours. The three bursts were observed within a 20-minute interval at phase $\sim 0.43$ of the active period, which peaks at phase 0.5. No other bursts were detected at P-band during the remainder of the campaign and no bursts at all were detected throughout the whole campaign at L-band. 
Given our nominal sensitivity, we expect that we would have been able to detect at least three of the four bursts detected by Effelsberg at L-band in \cite{marcote20}.

Assuming a Poissonian distribution of the events in the $\pm2.6$ days of activity of the source, and with the hypothesis that our sensitivity did not change during the time span of the observations, except for little variation of the RFI environment, we calculated the probability that our detections all happened during the first hour of observations. If we assume the rate of $\sim 1$~burst~h$^{-1}$ from CF20, we obtain a probability $P(3)=4\times10^{-10}$ that the SRT should observe only three bursts during the full time span of its observations, and obviously the probability is even lower for the 3 bursts to all occur within a 20 min interval. While it is true that no fluence completeness distribution can be derived for the SRT at the moment, and that the two instruments have different sensitivity limits, the aforementioned probabilities seem to discard a uniform distribution of bursts at 328 MHz along the $\sim 5.2$ day window; they favor a clustering of the bursting activity of \frb,  as also hinted at by CHIME observations. \cite{zanazzi_lai20} propose that this clustering is expected in the framework of their precessing magnetar model. This is because the viewing angle of an observer constantly changes phase with respect to the neutron star's inclination angle due to precession, and the intensity of the bursts changes accordingly.


\subsubsection{Scattering} \label{subsec:radioscattering}

\begin{figure}
    \centering
    \includegraphics[width=0.5\textwidth]{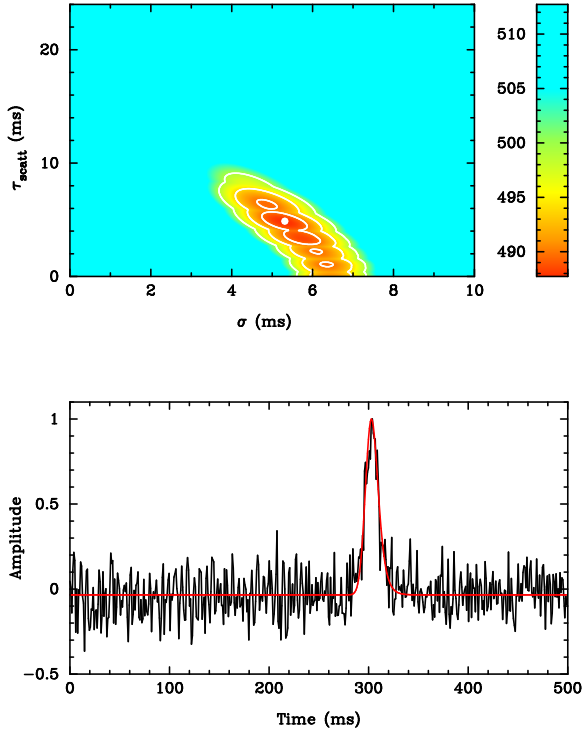}
    \caption{Upper panel -- Left: $\sigma_{G}-\tau_S$ $\chi^2$ color map for burst-1; the white dot indicates the $\sigma_{G}-\tau_S$ couple of values for which $\chi^2$ is minimum; the white lines trace the contour levels at 1-, 2- and 3-$\sigma$, moving away from the white dot, respectively. Right: reference color scale for the $\chi^2$ color map. 
    Lower panel -- Burst-1 profile resolved in 512 time bins along the displayed 0.5 seconds of data. The red superposed line is the simulated scattered profile that results from the couple of values $\sigma_{G}-\tau_S $ for which the $\chi^{2}$ is minimum. Both profiles have been normalized to have zero average and peak amplitude 1.}
    \label{fig:scatt}
\end{figure}

We investigated the possible presence of a scattering tail in the observed profiles, by comparing them to simulated scattered ones.
The code we used to generate the simulated profiles was designed for radio pulsar profile investigations, and requires, as a reference, an observed profile at a frequency that is high enough to be unaffected by this phenomenon.
In this case, given the lack of such a reference profile, we assumed that the unscattered profile is well approximated by a Gaussian curve. 
We then exploited the 2D $\sigma_{G}-\tau_S$ space, where $\sigma_{G}$ is the unknown width of the unscattered Gaussian, and $\tau_S$ is the scattering time at our observing frequencies.
We compared the simulated profile corresponding to each couple $\sigma_{G}-\tau_S$ to the observed one, and assigned a $\chi^2$ value to each of them:

\begin{equation}
    \chi^2(\sigma_{G},\tau_S,\phi_{0})=
    \int_0^1[S(\sigma_{G},\tau_S,\phi-\phi_0)-P(\phi)]^2d\phi
\end{equation}
where $S(\sigma_{G},\tau_S,\phi)$ is the simulated profile, $P(\phi)$ is the observed one, and $\phi_0$ is the phase shift between the two profiles for which $\chi^2(\sigma_{G},\tau_S,\phi_{0})$ is minimum. 
The upper panel of Fig.\,\ref{fig:scatt} displays the resulting $\chi^2$ map for burst\,1 along with the 1-, 2- and 3-$\sigma$ contour levels. The lower panel shows the observed profile resolved in 512 bins along the displayed 0.5 seconds of data, with, superimposed, the simulated scattered profile that results from the couple of values $\sigma_{G}-\tau_S$ for which the $\chi^2$ is minimum. The resulting $\chi^2$ maps for each of the 3 bursts indicate that the 2$\sigma$ ULs for $\tau_S$ at $328$\,MHz are 9.5\,ms, 10.2\,ms and 10.8\,ms for bursts 1, 2 and 3, respectively.

Once re-scaled to the reference frequency of 1\,GHz, they imply $\tau_S\leq0.1$\,ms, under the hypothesis $\tau_S\propto\nu^{-4}$ (where $\nu$ is the radio frequency). 
 This result fits in nicely
with the DM$-\tau_S$ correlation f reported by \cite{chime2019allfrb}, in their Fig. 2. 

It is worth noting that the inspection of the $\sigma_{G}-\tau_S$ $\chi^2$ maps shows that the three SRT bursts are also compatible with being immune to (or having a very low level of) scattering, which means that the detection at frequencies that are even lower than the SRT P band might be possible. For instance, a value of $\tau_S\sim 1$\,ms at $328$\,MHz would translate to modest scattering delays ($\sim 10-20$\,ms) at frequencies around $150-190$\,MHz, which are typical of the LOFAR High band \citep{vanHaarlem+13,stappers}. On the other hand, for $\tau_S$ values approaching the ULs reported above, the burst energy would be diluted over $0.8-1.0$\,s, easily causing the non-detection of similar bursts with LOFAR \citep{houben19}. Therefore, additional future observations of the periodic \frb\ at P-band (where the feasibility is now warranted) are needed to collect events that are bright enough to finally constrain the value of $\tau_S$. This is particularly crucial for assessing the role played in FRB science -- at least for nearby sources -- by future low frequency telescopes, such as SKA-LOW, which will be operating in the 50--350 MHz band.

\subsection{Spectral properties of the radio bursts} \label{subsec:radiobursts}

Several explanations were proposed to explain the lack of detection of FRBs or RFRBs at low frequencies.
\citet{ravi19a} suggested that free-free absorption by electrons in the intervening medium or, alternatively, induced Compton scattering can be responsible for the non-detections below 400 MHz \citep{chawla,sokolowski,rajwade20a}. \citet{rajwade20a}, in particular, showed that induced Compton scattering alone cannot account for the lack of detections in their 332~MHz FRB survey. They suggest that free-free absorption should play a more prominent role, most likely happening in post-shock regions of super-luminal supernovae (SNe) where the electron densities can reach $n_e \sim 10^5$~cm$^3$ \citep{margalit18}. \citet{sokolowski} reached similar conclusions, although without discriminating between physical absorption models. The FRB~180916 case is a peculiar example of the opposite scenario, where the three bursts are seen at 328~MHz but are undetected at 1.4~GHz, revealing the absence of a spectral turnover.

Barring any specific modeling, the detection of three bursts at 328 MHz confirms that (R)FRB environments can be optically thin to this emission and that a cutoff for the bursting mechanism, if present, must be at even lower frequencies.
Although FRBs have not yet shown simultaneous emission over a wide radio frequency band, simultaneous radio observations in two widely separate bands - as shown here - can be extremely valuable for constraining the instantaneous apparent spectral properties of the bursts. In fact, it is worth noting that the observed spectrum of FRBs is likely significantly affected by the effects of the medium the signal went across (see e.g. \citealt{cc19,cordes17}) and hence might not map the intrinsic spectral properties of the source.

Our single-frequency detection does not allow us to place constraints on any specific physical mechanisms, but our detection threshold is comparable to \citet{rajwade20}, indicating that the environment of a super-luminal supernova (SN) seems unlikely for this source - even though it is located in a star-forming region \citep{marcote20}.  The absence of a bright SN is consistent with the fact that \cite{marcote20} do not find accompanying persistent emission to \frb\ down to a limiting sensitivity that is 400 times lower than the detected emission of FRB~121102.
In particular, the non-detection at L-band allows us to set an UL  on the burst spectral index\footnote{We use the convention $F_\nu \propto \nu^{-\alpha}$, where $F_\nu$ is the flux density at the frequency $\nu$.} $\alpha \sim 1$ for the brightest burst, assuming the nominal L-band sensitivity.
 A modulation due to plasma lenses \citep{cordes17} remains an open  possibility, although the lack of simultaneous detection at 1.4~GHz makes it not obvious; a future, dedicated analysis will be required.

\subsection{Burst energetics\label{subsect:energetic}}
The total (assumed isotropic) energy $E_{r,i}$ emitted in the observed radio band during a radio burst can be approximated as: 
\begin{equation}
E_{r,i} \simeq  2.7\,\times\, 10^{37} \, S_{\nu, Jy} \, \delta t \, \Delta \nu \, d_{150M}^2 \, \, {\rm erg},    
\label{eq:radio_energy}
\end{equation}

where $S_{\nu}$ is the measured FRB peak flux density in Jy, $\delta t$ is the duration of the burst in ms, $\Delta \nu$ is the bandwidth in GHz and $\rm d_{150M} = \, d/150\, \rm Mpc$  with $d$  the source distance from Earth. For burst-1 from SRT, $S_{\nu} = 2.8$~Jy, $\delta t = 13$~ms, and $\Delta \nu = 0.064$~GHz, resulting in $E_{r,i} \simeq 6.3\,\times\, 10^{37}\, {\rm erg}$. 

Where does this energy output come from? The plethora of proposed theoretical models about the emission mechanism(s) -- and hence the energetics -- of Repeating FRBs (RFRBs) mainly rely on the observed properties of the original source, FRB~121102. In particular, most hypotheses invoked a relatively young magnetar spinning at a millisecond period, whose radio bursting activity might be powered by its high magnetic field and resulting from either internal processes or triggered by the interaction with an extreme magneto-ionic environment (see e.g. \citealt{beloborodov17,margalit18}). 

As reported in \S 1, similar considerations could be applied to the case of FRB~180916, once suitably adjusted to explain its periodicity. However, there are also significant differences between the properties of the two RFRBs (e.g. the nature of the host galaxy, \citealt{tendulkar17,chatterjee17,marcote17,marcote20}, the Rotation Measure, \citealt{michilli18,gajjar18}, CF20), which may lead us to explore other possibilities, such as the idea (see e.g. \citealt{corwas16}) of a predominant role of the spin-down power of a neutron star (NS) in shaping the energetics of FRB~180916. In fact, emission from pulsar giant pulses (considered as a manifestation of the conversion of rotational energy into coherent bright radio emission), as observed for instance from the Crab pulsar (see e.g. \citealt{mickaliger12,haneil07}), was proposed by some authors \citep{lyutikov16} as a favored emission mechanism for the FRB, as compared to a magnetic-powered scenario, on the basis of both the energetics and other properties of the pulsar giant pulses (i.e. polarization, spectra) that are reminiscent of what is seen in FRBs. 

The energy of burst-1 is several orders of magnitude above the typical energy of the Crab's giant pulses (see e.g. \citealt{mickaliger12,hankins03}). However, it was noted \citep{corwas16} that ms-duration emission can be the result of the coherent addition of ns-shot emission, similarly to the strongest examples observed in the Crab pulsar, in which unresolved 0.4 ns bursts of $S_{peak}\sim 2$\,MJy have been detected (\citealt{haneil07}). \cite{corwas16} suggested that a much longer monitoring of the Crab pulsar with respect to that available so far could show the occurrence of ultra giant pulses with energetics comparable to that of FRBs. In addition, much brighter pulses could be released by a NS that is younger and more energetic than the Crab. 

This hypothesis would in turn call for the \frb's source to be enshrouded in a SN remnant. \cite{piro16} studied this case and noted that the reverse shock, produced by the SN ejecta when they encounter the interstellar medium, generates free electrons that can disperse and absorb the radio signal. Hence, emission down to $\sim 300$\,MHz can escape from free-free absorption only when the intervening matter is diluted enough, i.e. at least $300-500$ years after the SN explosion, depending on the density of the medium and, mostly, on the mass of the ejecta. Following this model (Fig. 5 of \citealt{piro16}), and assuming the usually invoked formulae relating $P$ and $\dot P$ to the age and the surface magnetic field of a neutron star (see e.g. \citealt{lk04}), the energy released by burst-1 could originate from the spin-down power of a NS of age larger than about $300$\,yrs, spin period of a few ms, and surface magnetic field $B \leq 10^{11}$\,G. This picture could be validated by the observation of a progressively-decreasing dispersion measure associated with the pulses during the next few years.

Very recent developments in the field are showing that a border land may be connecting the two aforementioned energy sources for the RFRBs, which might also be simultaneously at work. In March 2020, \cite{esposito20} observed the just discovered \citep{enoto20}, and possibly also the youngest \citep{champion20}, Galactic magnetar, Swift-J1818.0--1607. Seven days after the occurrence of an X-ray burst -- which is usually interpreted as a magnetic-powered event --, an observation with SRT showed that the radio emission from Swift-J1818.0--1607 (in turn discovered shortly after the X-ray burst \citealt{karuppusamy20}), was at that time predominantly occurring in the form of strong and sporadic radio pulses. In April 2020, the already known Galactic magnetar SGR J1935+2154 was observed by Swift \citep{palmer20}, soon followed by many X-ray instruments, to produce a forest of X-ray bursts. Four days later, \cite{scholz20b} reported a single ms-duration radio burst observed by CHIME and coming from the same source with an estimated fluence of $\sim $\,kJy~ms; it was also independently detected by STARE2 \citep{stare2} at L-band, with an estimated fluence $>1.5$~MJy~ms. Despite extensive radio follow-ups, at the date of writing, only one other burst has been reported since then: it was observed by FAST six days after the original X-ray event, with an estimated fluence of 60~mJy~ms at L-band \citep{zhang20atel}. While these radio bursts, from an energetics point of view, are all still not bright enough to match the more energetic FRBs, this is the closest example we have come so far to directly associate a class of known sources (magnetars and pulsars), as well as of emission properties (X-ray bursts and very strong radio pulses), to a phenomenology resembling that of RFRBs.

\begin{figure}
    \centering
    \includegraphics[width=1.\columnwidth]{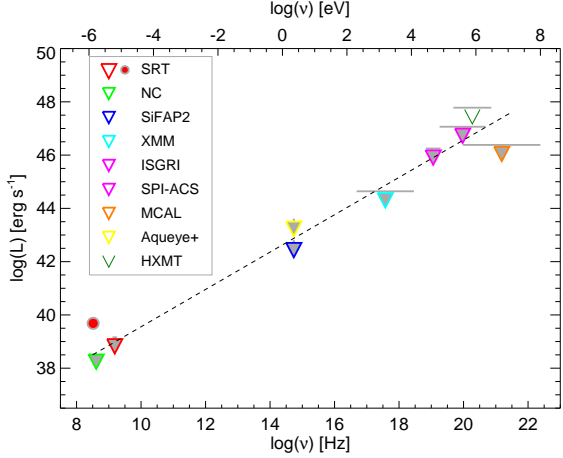}
    \caption{Burst luminosity vs frequency diagram throughout the electromagnetic spectrum for the observations performed during our multiwavelength campaign. The red dot represents the SRT burst-1 detection; the remaining grey-filled triangles represent ULs obtained on the ms-fluence by the instruments involved in our campaign that can have sensitivity in the ms range. SRT L-band observation, {\it XMM-Newton} and {\it AGILE-MCAL} were simultaneous with the detection of burst-1, while the other grey-filled triangles represent the sensitivity limits probed by the instruments that do not have simultaneous observations with SRT bursts.  SRT L-band data and the NC data refer to a 13-ms burst and the ULs are calculated at the $6 \sigma$ level, as it is standard for the radio band, as opposed to the $3 \sigma$ ULs indicated for the multi-wavelength counterparts.   INTEGRAL SPI-ACS data refer to a 50-ms limit, as specified in the text (\S \ref{subsec:resultx}). The dark-green, unfilled, triangle represents the corresponding UL reached by HXMT using archival data \protect\citep{guidorzi20}. The black dashed line with slope 0.70 is not a fit and is only to be intended as a guide for the eye. }
    \label{fig:lum_vs_nu}
\end{figure}

\subsection{The multi-wavelength campaign and constraints to models} \label{subsec:radioburstsmodel}
Fig. \ref{fig:lum_vs_nu} shows a summary of the ms-fluence ULs obtained using the instruments involved in our campaign.
{\it XMM-Newton} and AGILE were active at the time of the detected bursts and therefore their ULs, together with SRT's L-band, constitute a punctual limit on the high energy emission from the SRT bursts. The ULs displayed for the NC and the optical telescopes are, on the other hand, an estimate of the sensitivity reached by our survey. For comparison, we show HXMT UL that was recently published in \cite{guidorzi20}.

Simultaneous X-ray observations with {\it XMM-Newton} in the X-ray range led to upper limits on the persistent and burst luminosities, of $\sim10^{41}$ and $\sim10^{45}$~erg\,s$^{-1}$, respectively (under the assumptions for the spectral shapes reported in \S\ref{subsec:resultx}). In the case of an isolated young pulsar scenario, this persistent limit is not constraining, given that the typical X-ray luminosity of a young rotational powered pulsar or a magnetar are $<10^{38}$~erg\,s$^{-1}$. As for the bursting luminosity, we can compare the ULs related to the available XMM, INTEGRAL, and AGILE data (see below) to check the level at which ultra energetic events can be constrained.

X-ray and hard X-ray observations with INTEGRAL (not simultaneous with the SRT radio bursts) on 0.1--1 sec timescales provide ULs to the fluence in the range $(0.6 - 2) \times  10^{-8} \rm \, erg \, cm^{-2} \,  s^{-1}$, and therefore an UL on the total energy in the range $(1 - 4) \times 10^{46} \rm \, erg $.

As for AGILE, we focus on burst-1, which has the largest radio fluence among those detected by SRT. \agilep/MCAL provides an interesting UL in the MeV range simultaneously with this radio detection. The value of $F'$ translates into an UL for the isotropically radiated energy into the MeV range, $ E_{MeV,UL} =\, 4 \, \pi F' \, d_{150M}^2 \simeq (2.2 \,\times\, 10^{46} \, \rm erg) \, d_{150M}^2$. This value of $ E_{MeV,UL} $ should be compared with the energy emitted in the radio band, $E_{r,i} \simeq 6.3\,\times\, 10^{37}\, {\rm erg}$.  Therefore, our UL on the energy emitted in the MeV range on millisecond timescales is less than $\sim 5 \times 10^8$ times the energy emitted at 328 MHz for this radio burst.

It is interesting to compare the various high-energy ULs reported above with the energy emitted by the soft gamma-ray repeater SGR 1806--20 (e.g. \citealt{turolla2015}) during its Giant Outburst in 2004, which lasted about 200 ms {\citep[][]{Palmer2005,Hurley2005}}, while burst-1 lasted $\sim\, 20\, $ms. In that case, an isotropic energy  $ E_{MeV} \sim (2 \,\times\, 10^{46} \, \rm erg)$ was emitted in the MeV range. We can thus exclude, as a counterpart to burst-1 in X-rays, a Giant Burst that is at least twice as bright as SGR 1806--20. As already mentioned in \S \ref{subsect:energetic},
 Integral, AGILE, Konus-Wind and HXMT \citep{mereghetti20,tavani20b,konus,hxmt} most recently detected the hard X-ray counterpart to the radio burst from SGR~J1935+2154 observed by CHIME and STARE2. The preliminary estimated fluence of the high energy burst, compared to the radio one, indicates that current telescope sensitivities might still be too high to detect high energy emission even from relatively close FRBs such as \frb.

 During the reviewing stages of this manuscript, the results of a multi-wavelength campaign involving the {\it Chandra} X-ray telescope observing \frb\ simultaneously with CHIME were reported by \cite{scholz20}. Their conclusions agree with our findings.

\section{Summary} \label{sec:sum}

Our multi-wavelength observations of FRB 180916 provide a wealth of valuable information on this puzzling source. The SRT detection of three strong radio bursts at 328 MHz represents the first firm detection of any FRB type at such low frequencies. It also shows that no significant scattering is affecting the emission at these frequencies. On the one hand, this confirms that the scattering time and DM seem to correlate \citep{chime_rep19,cc19}; on the other hand, this leaves open a detection of (at least relatively nearby) FRBs at even lower frequencies with instruments such as LOFAR and SKA-LOW.
The lack of a simultaneous detection at 1.5 GHz is also relevant: it confirms the narrow-band emission typically seen from Repeaters, indicating either a relatively steep spectrum of the intrinsic radio emission or a strong effect of the intervening medium.

The source is capable of emitting $\sim 10^{38} \rm \, erg$ within a few tens of milliseconds at 328 MHz, that is an energy many orders of 
magnitude larger than the giant pulses from Crab-like pulsars. In addition, the upper limits resulting from our simultaneous observation in the keV--MeV range for the first radio burst detected by the SRT are still compatible with a high-energy activity that is similar to the 2004 giant outburst of SGR 1806--20. Additional upper limits obtained from our non-simultaneous observations in the optical, X-ray, hard X-ray and gamma-ray energy ranges constrain the emission of the FRB 180916 source to be less than $10^6 - 10^8$ times the Eddington luminosity for a one solar-mass compact object.

The periodicity of the bursting activity of \frb\, represents an unprecedented observational opportunity to characterize its properties. Given the erratic behavior - from the optical to the high-energy frequencies - that is expected in the various proposed models, the long-term radio monitoring at low and intermediate frequencies, as well as deep observations with the best available instruments across the electromagnetic spectrum, might finally allow us to determine the nature of the source and strongly constrain the mechanism of burst production.

\section*{ACKNOWLEDGMENTS}
The authors thank the anonymous referee for helpful suggestions which helped improve the paper. M.P. wishes to thank Marilyn Cruces and Dongzi Li for useful discussions.
A.Po., M.B. and A.R. gratefully acknowledge financial support from the research grant ``iPeska'' (P.I. Andrea Possenti) funded under the INAF national call PRIN-SKA/CTA with Presidential Decree 70/2016.
V.G. is supported by Breakthrough Listen, which is managed by the Breakthrough Initiatives and sponsored by the Breakthrough Prize Foundation (http://www.breakthroughinitiatives.org).
F.P.  acknowledges financial support from ASI/INAF agreement n.2019-35-HH.0 and PRIN-INAF SKA-CTA 2016. 
A.Pa. and L.Z. acknowledge financial support from the Italian Space Agency (ASI) and National Institute for Astrophysics (INAF) under agreements ASI-INAF I/037/12/0 and ASI-INAF n.2017- 14-H.0 and from INAF ‘Sostegno alla ricerca scientifica main streams dell’INAF’, Presidential Decree 43/2018.
N.R. is supported by the ERC Consolidator Grant ``MAGNESIA" under grant agreement Nr. 817661, Catalan grant SGR2017-1383, Spanish grant PGC2018-095512-BI00, and acknowledge support from the PHAROS COST Action (CA16214).
The Sardinia Radio Telescope is funded by the Department of Universities and Research (MIUR), the Italian Space Agency (ASI), and the Autonomous Region of Sardinia (RAS), and is operated as a National Facility by the National Institute for Astrophysics (INAF). We thank the SRT staff for the timely scheduling of the observations.
We thank Norbert Schartel for approving {\it XMM-Newton} Directors Discretionary Time and the {\it XMM-Newton} SOC staff for scheduling the observations.
The SiFAP2 team thanks the TNG Director Ennio Poretti for the time scheduled within the time reserved for technical purposes. 
The results presented in this paper are based on the observations made with the Italian {\it Telescopio Nazionale Galileo} (TNG) operated by the {\it Fundaci\'on Galileo Galilei} (FGG) of the {\it Istituto Nazionale di Astrofisica} (INAF) at the {\it  Observatorio del Roque de los Muchachos} (La Palma, Canary Islands, Spain) and on observations collected at the Copernicus and Schmidt telescopes (Asiago, Italy) of the INAF-Osservatorio Astronomico di Padova and at the Galileo telescope (Asiago, Italy) of the University of Padova.

\facilities{SRT, Northern Cross, XMM-Newton, NICER, INTEGRAL, AGILE, TNG, Asiago}

\bibliography{biblio,References}
\bibliographystyle{aasjournal}

\end{document}